\documentclass{KapProc} 

\usepackage{epsf}
\usepackage{graphicx}
\normallatexbib

\newcommand{\ep}{\varepsilon}
\renewcommand{\k}{{\bf k}}

\begin{document}

\articletitle{Generalized conductance sum\\ rule in atomic break junctions}


\author{S.~Kirchner$^a$, J.~Kroha$^a$, and E.~Scheer$^b$} 
\affil{$^a$ Institut f\"ur Theorie der Kondensierten Materie, 
Universit\"at Karlsruhe\\ \hspace*{0.3cm}D-76128 Karlsruhe, Germany\\
$^b$ Universit\"at Konstanz, Fachbereich Physik,
D-78457 Konstanz, Germany}

\email{kroha@tkm.physik.uni-karlsruhe.de}

\chaptitlerunninghead{Generalized conductance sum rule in atomic break junctions}


\begin{abstract}
When an atomic-size break junction is mechanically stretched, 
the total conductance of the contact remains approximately constant
over a wide range of elongations, although at the same time 
the transmissions of the individual channels (valence orbitals of the 
junction atom) undergo strong variations. 
We propose a microscopic explanation of this phenomenon, based on
Coulomb correlation effects between electrons
in valence orbitals of the junction atom. 
The resulting approximate conductance quantization is closely related
to the Friedel sum rule. 
\end{abstract}

\begin{figure}
\centerline{\includegraphics[height=5.0cm]{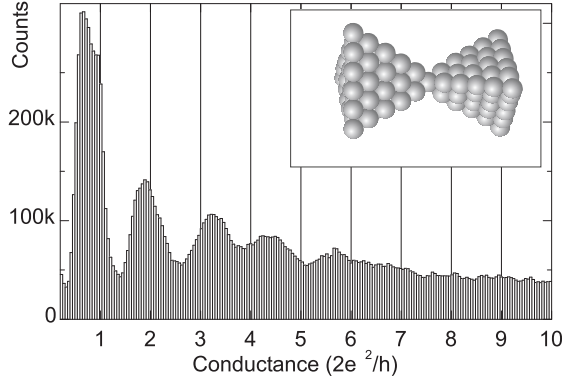}}
\caption{Histogram of the frequency of occurance of various conductance
values of an Al atomic break junction during $> 6200$ opening 
sweeps of the contact; $T=4.2$~K. 
The occurance of preferred conductance values
close to multiples of the conductance quantum $2e^2/h$ is seen 
(courtesy: A.~I.~Yanson and J.~M.~v.~Ruitenbeek \cite{ruit.97}). The inset 
schematically shows the junction geometry.}
\label{fig:histo}
\end{figure}

The progressing miniaturization of electronic circuits raises the question
what controls the transport when a contact is shrunk to its minimal 
possible size, a single atom. Electrical single-atom contacts have
recently been fabricated using the break junction technique \cite{scheer.97}. 
By analyzing the subgap structure of superconducting contacts 
it was demonstrated that the valence orbitals of the junction atom act as
the transmission channels for the electronic current and that the
transmissions $T_m$ of the individual orbitals $m=1,\dots, N$, add up
to the total transmission of the contact, which was measured independently
\cite{scheer.97}.
Naturally, the transmissions $T_m$ can take any value $0\leq T_m \leq 1$,
since they depend on the microscopic details like the coupling matrix
elements of the atomic orbitals to the leads. Therefore, it came as 
a surprise that in Al junctions the total 
conductance remained nearly constant with a value close to the 
conductance quantum $2e^2/h$, 
when the contact was mechanically stretched, although 
at the same time the individual channel transmissions $T_m$ varied over a
wide range \cite{ruit.97}. As a consequence, certain 
conductance values are preferred in Al contacts 
as shown in Fig.~\ref{fig:histo}. 
In the present work we prove a sum rule for the total conductance
in atomic junctions where the Coulomb repulsion between electrons
in the valence orbitals of the junction atom is large. We propose that
this correlation effect is the microscopic origin of the observed
approximate conductance quantization. 

In good metals the Coulomb interaction is screened on a length scale
of a few \AA. For electrons in the extended conduction band states of the leads
it can, thus, be absorbed into a small renormalization of the
Fermi liquid parameters. However, electrons traversing the contact
are forced to pass through the orbitals of the junction atom,
where their dynamics are strongly affected by the 
Coulomb electron-electron interaction because of the strong spatial 
confinement. As a model of the junction we, therefore,
consider the Anderson-like Hamiltonian
\begin{eqnarray}
\label{eq:hamilton}
  H &=& \sum\limits_{k\sigma,\;\alpha=L,R} \ep_{k}\; 
                          c^{\dagger}_{\alpha\;k\sigma}
                          c^{\phantom{\dagger}}_{\alpha\;k\sigma} +
 \sum\limits_{m\sigma} \ep_{d,m}\; d^\dagger_{m\sigma}
                                  d^{\phantom{\dagger}}_{m\sigma} \\
  &+& \sum\limits_{k m\sigma ,\; \alpha = L,R} 
    \left[ t^{\alpha}_{mk}\;d^\dagger_{m\sigma} 
           c^{\phantom{\dagger}}_{\alpha\; k\sigma} 
    +{\rm h.c.}\right] 
  + \frac{U}{2} \sum\limits_{(m,\sigma)\neq(m',\sigma ')} 
        \hat n_{m\sigma} \hat n_{m' \sigma '} \; , \nonumber
\end{eqnarray}
where $c^{\dagger}_{\alpha\;k\sigma}$ is the creation operator of
an electron in the left or right lead ($\alpha = L,R$) with energy $\ep _k$ 
and spin $\sigma$. $d^\dagger_{m\sigma}$ creates an electron in one of
the valence levels $\ep _m$, $m=1,\dots ,N$ of the junction atom, and
$\hat n_{m\sigma} = d^\dagger_{m\sigma} d^{\phantom{\dagger}}_{m\sigma}$.
The transition matrix elements from a lead state $(\alpha ,k)$ to a local 
level $m$
is $t^{\alpha}_{mk}$, and the Coulomb repulsion between two electrons in
any of the junction atom's valence orbitals is denoted by $U$.
To be explicit, we restrict ourselves to the case of $N=2$ transmission
channels here (A similar conductance sum rule can be proven for arbitrary
$N$ \cite{kroha}). 
The advanced local-orbital Green's function then takes the  
matrix form 
\begin{eqnarray}
{\cal G}_{\sigma}(\omega ) =  \left( \begin{array}{cc}
\omega-\ep_{d,1}-i\Gamma_{11}-\Sigma_{1}(\omega) &     -i\Gamma _{12} \\
 -i\Gamma _{21}         &   \omega-\ep_{d,2}-i\Gamma_{22}-\Sigma_{2}(\omega)  
                   \end{array}
                   \right) ^{-1}
\label{eq:green}
\end{eqnarray}
with the effective lead-to-orbital coupling matrix $\Gamma = (\Gamma _{mn})$,
$\Gamma _{mn} = 
\pi \sum _{k,\;\alpha} t_{mk}^{\alpha} A_k(\omega ) t_{kn}^{\alpha\;*}$,\
($A_k(\omega )$: spectral density of lead state $k$).
$\Sigma _m (\omega ) = \Sigma _m '(\omega ) + i\Sigma _m ''(\omega )$
denotes the advanced self-energy due to Coul\-omb interaction $U$ in the local
orbitals. When a bias voltage $V$ is applied, the current is given 
by \cite{wingreen.92}
\begin{eqnarray}
I=\frac{e}{h} \sum _{\sigma} \int d\omega 
\Bigl[f(\omega )-
      f(\omega +\frac{eV}{\hbar} )\Bigr] {\rm Im}\, {\rm tr}\,   
(\Gamma \cdot {\cal G}_{\sigma}(\omega )) ,
\label{eq:current}
\end{eqnarray}
where $f(\omega )$ is the Fermi function. The hybridization $t^{\alpha}_{nk}$
induces an antiferromagnetic spin exchange coupling between
an electron in any of the atomic orbitals and the conduction electrons.
It is known from a renormalization group analysis of this problem, that 
the ground state of correlated quantum impurity models like 
Eq.~(\ref{eq:hamilton}) is a spin singlet \cite{hewson.93}. 
Hence, for temperature $T$ below a characteristic scale, $T<T_o$ 
the junction is a pure potential scatterer
for electrons traversing the system, and the following Fermi liquid
relations hold \cite{luttinger.60}, 
\begin{eqnarray}
\label{eq:fl}
&\phantom{=}&
\Sigma ''_m (\omega ) = [(\hbar\omega) ^2 + (\pi\k_BT)^2]/k_BT_o 
\hspace*{0.65cm}\omega, T \to 0\\
&\phantom{=}&
\int _{-\infty}^{0} d\omega\; {\rm tr}\Biggl\{ 
   \frac{\partial \Sigma (\omega)}{\partial \omega} \cdot 
   {\cal G}_{\sigma}(\omega) \Biggr\}=0
\hspace*{1cm} \mbox{(Luttinger theorem)}\hspace*{1cm}
\label{eq:lutt}
\end{eqnarray}
The averaged total electron number on the junction atom for each
spin species, $n_{d,\sigma}$, 
can now be evaluated using the general relation   
$\frac{d}{d\omega}{\rm ln}({\cal G}^{-1}) = 
(1 - \frac{d\Sigma}{d\omega})\cdot {\cal G}$
and the Luttinger theorem Eq.~(\ref{eq:lutt}),
\begin{eqnarray}
n_{d\sigma} = {\rm Im} \int _{-\infty}^0 \frac{d\omega}{\pi}
{\rm tr}\; {\cal G_{\sigma}(\omega )} =
\frac{1}{\pi} {\rm Im} 
\Bigl[{\rm tr}\{ {\rm ln}\; {\cal G}_{\sigma}(\omega )^{-1}\} 
\Bigr] _{\omega =-\infty}^{\omega = 0}\; .
\label{eq:friedel}
\end{eqnarray}
Eq.~(\ref{eq:friedel}) is a statement of the Friedel sum rule
$n_{d\sigma} = \frac{1}{\pi} \sum _{m} \delta _{m\sigma} (0) $, since 
the scattering phase shift at the Fermi level in channel $m$
is $\delta _{m\sigma}(0) =
{\rm arg} [\Gamma \cdot {\cal G}_{\sigma}(0)]_{mm}$.
It may be re-expressed, using 
${\rm tr}\;{\rm ln}\;{\cal G_{\sigma}}^{-1} = 
{\rm ln}\;{\det}\;{\cal G_{\sigma}}^{-1}$, as
\begin{eqnarray}
n_{d\sigma} =  \frac{1}{\pi}{\rm arccot}
 \Biggl[ \frac{{\rm Re}\;{\rm det}\;{\cal G}_{\sigma}(0)^{-1}}
              {{\rm Im}\;{\rm det}\;{\cal G}_{\sigma}(0)^{-1}}\Biggr]\; .
\end{eqnarray}
The scattering T-matrix of the junction atom, 
$\Gamma \cdot {\cal G} _\sigma$, which determines the conductance $G=dI/dV$
of the system via Eq.~\ref{eq:current}, is now evaluated
by expressing the inverse matrix Eq.~(\ref{eq:green}) 
in terms of its determinant, and, using the Fermi liquid property
Eq.~(\ref{eq:fl}), we obtain at the Fermi energy ($\omega =0$, $T\ll T_o$),
\begin{eqnarray}
\label{eq:unitarity}
{\rm Im}\; {\rm tr}\; (\Gamma \cdot {\cal G}_{\sigma}(0)) &=& 
{\rm sin}^2(\pi n_{d\sigma}) \\ 
&+& {\rm sin}(2\pi n_{d\sigma}) 
\frac{\Gamma_{21}\Gamma_{12}-\Gamma_{11}\Gamma_{22}}
{\Gamma_{11}(\ep_{d,2}+\Sigma _2'(0))+\Gamma_{22}(\ep_{d,1}+\Sigma _1'(0))}\; .
\nonumber
\end{eqnarray}
If the transition amplitudes $t^{\alpha}_{mk}$ are independent of the
lead channels $k$, it follows directly from the definition of $\Gamma _{mn}$
that the term $\propto {\rm sin}(2\pi n_{d\sigma})$ cancels. 
Eq.~(\ref{eq:unitarity}) is an exact result, valid for arbitrary
microscopic parameters $\Gamma _{mn}$, $\ep _{d,m}$, $U$, and $n_{d\sigma}$.
It is the generalization of the well-known unitarity rule of the single-level 
Anderson impurity problem to the case of several impurity levels \cite{kroha}.
In metals in the single-atom junction geometry 
(inset of Fig.~\ref{fig:histo}) \cite{cuevas.98}, there is
at least one of the local levels significantly below the Fermi level
($\ep _{d,m_o} < 0$, $|\ep _{d,m_o}| /\Gamma _{mn} < 1$). 
While in the non-interacting case the right-hand side of 
Eq.~(\ref{eq:unitarity}) can assume any value, a strong Coulomb
repulsion $U$ enforces $n_{d\sigma} \approx 1/2 $, implying
via Eqs.~(\ref{eq:unitarity}), (\ref{eq:current}) a conductance 
close to the conductance unit, i.e.~$dI/dV \approx 2 e^2 /h$
(the factor 2 reflects spin summation). 
The physical origin of this quantization is that in the regime of 
large $U$ charge fluctuations are suppressed and the low-energy
spin fluctuations induce a Kondo-like resonance at the Fermi energy,
(as seen from the resonant phase shift $\delta (0) = \pi /2$,
which is implied by $n_{d\sigma}\approx 1/2$ through Eq.~(\ref{eq:friedel}).
Resonant transmission through the impurity complex is equivalent to
unitary conductance per spin. This quantization holds for the total conductance
and is exact in the limit $n_{d\sigma}=1/2$, $T\ll T_o$. It
will be approximate for the realistic parameters of a break junction. 

We thank J.~C.~Cuevas, P.~W\"olfle 
and J.~M.~van Ruitenbeek for useful discussions. 
This work was supported by DFG through SFB 195.

\end{document}